\documentclass{article}
\usepackage{graphicx} 
\usepackage[a4paper,margin=1in]{geometry}
\usepackage{setspace}
\usepackage{amsmath, amssymb}
\usepackage{xcolor}
\usepackage{graphicx}
\usepackage[super, comma, sort&compress]{natbib}
\usepackage{hyperref}
\usepackage{titlesec}
\usepackage{booktabs}
\usepackage{float} 
\usepackage{subcaption}
\usepackage{tabularx}
\usepackage{tcolorbox}
\usepackage{listings}
\usepackage{verbatim}
\usepackage{multirow} 
\usepackage{makecell}
\usepackage[normalem]{ulem}
\usepackage{pdflscape}
\usepackage{afterpage} 
\usepackage{array} 
\usepackage{authblk}
\newcommand{\indep}{\perp \!\!\! \perp}

\title{Toward a practical handbook for choosing among causal inference methods in non-randomized studies with binary outcomes: A simulation study for applied researchers}

\author[1]{Adrián Aurensanz-Crespo\thanks{These authors contributed equally to the paper and are joint first authors.}\thanks{Corresponding author: \texttt{aaurensanz@unizar.es}.}}

\author[2,3]{Cristóbal M Rodríguez-Leal\protect\footnotemark[1]}

\author[2,4]{Rosario Susi}

\author[1]{Jorge Castillo-Mateo}

\author[1]{Jesús Asín}

\author[5]{José M Ramírez} 

\author[2,4]{Teresa Pérez}

\affil[1]{Department of Statistical Methods and IUMA, University of Zaragoza, Zaragoza, Spain}
\affil[2]{Faculty of Statistical Studies, UCM, Madrid, Spain.}
\affil[3]{Emergency Department, Henares University Hospital. Foundation for Biomedical Research and Innovation of the Infanta Sofía University Hospital and the Henares University Hospital, Coslada, Spain}
\affil[4]{Institute of Statistics and Data Science. UCM, Madrid, Spain}
\affil[5]{Department of Surgery, University of Zaragoza, Zaragoza, Spain.}

\date{}

\begin{document}

\maketitle

\begin{abstract}
    Applied researchers in biomedicine and related fields are often interested in estimating the causal effect of a treatment or intervention. Although randomized clinical trials are considered the gold standard for establishing causal effects, they are not always feasible, and real-world data may represent the only available source of evidence. In such settings, causal effects must be estimated using statistical methods applied to observational data. Over the last few decades, modern causal inference methods based on the potential outcomes framework have emerged as useful tools in this field. However, many such techniques exist, and their performance depends on factors such as sample size, the proportion of treated patients, the proportion of patients experiencing the outcome, the magnitude of the treatment effect, the target estimand, and potential violations of the fundamental assumptions of causal inference. Given the wide range of available methods, selecting an appropriate approach can be challenging for applied researchers. This study uses a large-scale simulation experiment to address this issue and provide researchers with a guide in the form of a handbook for a binary treatment and a binary outcome. Particularly, we test four popular statistical techniques: propensity score matching (full matching), inverse of the probability weighting, G-computation, and targeted maximum likelihood estimation. The proposed handbook is applied to two real-world datasets to assess its practical utility: one comprising vulnerable patients with mild COVID-19 ($n=534$ patients and more than 50\% treated), and another of patients undergoing colorectal surgery ($n=3635$ patients and about 20\% treated). 
\end{abstract}


\newpage
\section{Introduction}

Estimating the effect of a treatment on a clinically relevant outcome, such as mortality or hospitalization, is a key aspect of medicine and related fields. The gold standard for achieving this estimation is a randomized clinical trial, but there are frequent situations in which this is unfeasible. This may be due to ethical or economic issues, meaning that observational studies are sometimes the only way to estimate this effect. Additionally, post-randomization changes in randomized clinical trials may introduce confounding bias, causing the estimated effects to differ from the causal effect guaranteed by randomization. Nevertheless, such analyses can provide more realistic estimates for specific subpopulations or real-world settings, as in per-protocol analyses or pragmatic trials\cite{pragmatic_trial}. It is in these situations that causal inference techniques become very important, because in those settings assignation of the treatment is not random and confounding bias may arise. Only these statistical approaches that aim to rigorously adjust for confounding make possible to obtain valid causal conclusions from observational data or pragmatic trials\cite{Whatif}.

A wide range of statistical methods are available to adjust confounding bias in observational studies. Multiple regression represents the earliest approach, and since the 1980s, more specialized causal inference techniques have been developed\cite{Greenland1986, stuart2010, Smith2022}. Not all techniques are suitable for every situation. Understanding the properties and limitations of each method is essential for selecting the most appropriate approach in real-world applications and for improving the reliability and interpretation of causal conclusions. As an example, methods that rely on stratification of confounding covariates, such as multiple regression and propensity score matching (PSM), are not valid for adjusting time-dependent confounding\cite{Whatif, naimi2017}. In less complicated settings, also different statistical techniques perform in different ways\cite{luque2018}. It can depend on factors such as sample size, treatment prevalence and outcome prevalence, among other factors. Futhermore, it can be related to practical violation of any of the four fundamental assumptions that must be met to compute a valid causal estimate. The four assumptions are (see Table \ref{tab:notacion} for a summary of the notation):

\begin{itemize}
    \item \textbf{No interference}. The counterfactual outcome of an individual $i$ under a specific treatment value $A = a$ does not depend on the treatment assignments of other individual $j$.
    \begin{equation*}
        Y_i^a\indep Y_j^a, \quad \text{where } i\neq j.
    \end{equation*}
    \item \textbf{Consistency}. The treatment values under comparison correspond to well-defined interventions, which in turn correspond to the versions of treatment observed in the data. The consistency assumption has two main components:
    \begin{itemize}
    \item First, counterfactual outcomes must be clearly defined, as clearly as the treatment itself.
    \item Second, counterfactuals must be linked to the observed data, which should contain sufficient information about the different versions of the treatment.
    \end{itemize}
    \begin{equation*}
        \text{If } A_i=a, \quad \text{then } Y_i^A=Y_i^a=Y_i.
    \end{equation*}
    \item \textbf{Conditional exchangeability}. The conditional probability of receiving every value of treatment, though not decided by the investigators, depends only on measured covariates $L$.
    \begin{equation*}
       Y^a\indep A \mid L, \quad \text{for all } a.
    \end{equation*}
    \item \textbf{Positivity}. The probability of receiving every value of treatment conditional on the confounding covariates must be greater than 0.
    \begin{equation*}
        P(A=a \mid L)>0, \quad \text{for all } a.
    \end{equation*}
\end{itemize}

\begin{table}[ht]
    \centering
    \renewcommand{\arraystretch}{1.2} 
    \caption{Summary of the notation used for causal inference, including confounding covariates, treatment and potential outcomes.}
    \begin{tabular}{ll}
        \hline
        Notation & Definition\\
        \hline
        $A$     & Treatment or intervention \\
        $a$     & Specific value of treatment $A$ \\
        $Y$     & Outcome \\
        $L$     & Confounding covariates \\
        $A_i$   & Treatment of individual $i$ \\
        $Y_i$   & Outcome of individual $i$ \\
        $Y_i^a$ & Counterfactual outcome of individual $i$ under specific value of the treatment $a$ \\
        $Y_i^A$ & Counterfactual outcome of individual $i$ under the treatment \\
        \hline
    \end{tabular}
    \label{tab:notacion}
\end{table}

All these assumptions tend to be guaranteed by design, but only the last one can be empirically assessed. Common settings where positivity assumption is violated or nearly violated are\cite{Zhou2020}:
\begin{itemize}
    \item Data limitations. Due to certain covariates, some individuals cannot receive one of the treatment options or will almost always receive only one of them. There is often a conceptual problem with the study design, so patients may have a nearly deterministic treatment assignment. This can happen when there is a strong indication or counter-indication for each treatment. Example: An observational study comparing two antibiotics for treating pneumonia, one of which is a beta-lactam antibiotic. Patients who are allergic to beta-lactams will always receive the non-beta-lactam antibiotic.
    \item Small sample size.
    \item Misspecification of the propensity score (PS) model.
    \item Limited overlap in the covariate distribution of the two treatment groups. This also usually leads to extreme weights that produces biased and unstable estimates with large variances. Example: An observational study of the efficacy of the flu vaccine among healthy young medical staff and elderly patients with multiple comorbidities.
\end{itemize}

The selection of appropriate estimators in non-ideal settings has been a central concern since the foundational formalization of observational study design, and it remains a challenge to this day\cite{rosenbaum2010design}. Furthermore, this issue remains highly relevant,  with strong interest driven by growing demand for real-world evidence to complement knowledge arising from randomized trials \cite{dahabreh2024, hernan2016}. While G-methods, including inverse probability weighting (IPW), G-computation (G-comp), and targeted maximum likelihood estimation (TMLE), are widely accepted as a robust framework for estimating causal effects \cite{naimi2017}, there is little practical consensus on which method to prioritize. This selection dilemma is aggravated in scenarios where fundamental assumptions are nearly violated. Recent investigations have highlighted how critical the choice of estimator becomes under positivity constraints \cite{leger2022}. Furthermore, latest studies demonstrate that this debate remains active across a broad spectrum of medical disciplines, where a heterogeneous array of methodologies continues to be simultaneously applied in search of the gold standard, due to the absence of a standardized selection framework to guide researchers \cite{burgos2025}. The ongoing uncertainty surrounding the state of the art highlights the urgent need for updated, detailed guidance for applied researchers.

Moreover, estimation of causal effect of a treatment can vary depending on the considered target population\cite{Austin2023, Whatif, Pirracchio2016}. When the research team wants to make the comparison in the whole population, the average treatment effect (ATE) is the contrast of interest. It answers the question of what difference would have been observed if every individual had been treated with one version of the treatment, compared with a counterfactual world where everyone had been treated with the other version of the treatment. When the comparison is made only in patients who receive one of the treatments, the average treatment effect on the treated (ATT) is the contrast of interest. Also, both ATE and ATT can be measured as an absolute measure, such as risk difference (RD), or as a relative measure, such as odds ratio (OR), relative risk (RR), and, in survival analysis, hazard ratio (HR). Regarding relative measures, it is crucial to acknowledge their non-collapsibility property. That implies that the marginal OR differs mathematically from the conditional OR even in the absence of confounding, unlike collapsible measures such as the RD\cite{Whatif,Greifer2025}.

The main objective of this study is to offer practical guidance for selecting causal inference methodologies in applied observational research for binary treatments and binary outcomes. Our simulation framework is designed to mirror the heterogeneity of real-world data analyses, jointly varying key study characteristics such as sample size, treatment prevalence, outcome prevalence, treatment effect, degree of positivity assumption violation based on the propensity score overlap (PS-OV), and the target causal estimand. By conducting a systematic comparison of PS-based, outcome-model-based, and doubly robust approaches that target both ATE and ATT across this broad set of scenarios, we aim to characterize their performance under realistic conditions and to develop a principled decision framework to help investigators to identify the most suitable method given the structure of their data and whether the estimand of interest is the ATE or the ATT. Finally, we apply the resulting handbook to two real-world observational datasets to illustrate its practical utility. The first dataset includes $n=534$ patients with mild COVID-19, characterized by a high treatment rate (63\%) and a moderate (40\%) outcome prevalence. The second dataset comprises $n=3635$ patients undergoing colorectal surgery, representing a contrasting clinical scenario with a lower proportion of treated individuals (23\%) and a low (26\%) outcome prevalence. We believe that our findings have direct implications for studies in medicine, epidemiology, and related fields, where sample size and PS-OV are often limited\cite{cenzer2020, pirracchio2012}.

The outline of this article is as follows: Section~\ref{sec:2} describes the compared statistical techniques, the data generating process, and studied performance measures, presenting the design of experiments developed. Section~\ref{sec:3} presents results in different settings with different degrees of PS-OV, and we give general recommendations to applied researchers. Also, we get deeper in results in frequent situations with moderate PS-OV, with detailed recommendations taking into account sample size, proportion of treated patients and proportion of patients with the outcome. Section~\ref{sec:4} applies the statistical techniques to real world datasets, and finally, Section~\ref{sec:5} discusses the meaning of the results, their practical implications, and the limitations of the study. It concludes with suggested future lines of investigation.

\section{Methods}\label{sec:2}
\subsection{Statistical techniques for causal inference}

A wide array of statistical techniques exists to estimate causal effects. These methodologies are typically categorized based on their modeling approach: those that focus on modeling the treatment assignment mechanism, specifically estimating the PS, defined as the probability of receiving treatment given the covariates:
\begin{equation*}
    PS = P(A=1 \mid L),
\end{equation*}
known as PS-based methods\cite{pirracchio2012}. Those that focus on modeling the relationship between covariates and the outcome,
\begin{equation*}
    E(Y \mid A,L) = P(Y = 1 \mid A,L), 
\end{equation*}
known as outcome-based methods. Finally, hybrid approaches that combine both strategies (double robust methods). Within this landscape, the G-methods\cite{naimi2017} provide a robust framework that generalizes these approaches, exploiting conditional exchangeability given a vector of confounders $L$ to estimate the causal effect of $A$ on $Y$ in the entire population or specific subsets\cite{Whatif, Smith2022, Marginal_Effects}. The methodologies compared in this article are:

\begin{itemize}
	
    \item IPW. This method uses the estimated PS to assign a weight to each observation, creating a balanced pseudopopulation in which treatment assignment is independent of measured confounders. It allows estimating causal effects via a simple weighted mean or regression. We implemented the normalized (or Hajek) estimator to improve finite-sample stability. Its expectations are defined in Equation \ref{eq:ipw1}.

    \begin{equation}
    \label{eq:ipw1}
    \hat{E}(Y^1) = \frac{\sum_{i=1}^n \hat{w}_{1i} A_i Y_i}{\sum_{i=1}^n \hat{w}_{1i}A_i} 
    \quad \text{and} \quad
    \hat{E}(Y^0) = \frac{\sum_{i=1}^n \hat{w}_{0i} (1 - A_i) Y_i}{\sum_{i=1}^n \hat{w}_{0i}(1 - A_i)},
    \end{equation}
    where $\hat{w}_{1i}$ and $\hat{w}_{0i}$ are the weights estimated to the individual $i$, which depend on whether or not they received the treatment, respectively. This weighting approach allows for calculating different estimators, depending on how weights are defined\cite{MathchIt, Zhou2020, Zhou2022}:  
    
    \begin{equation*}
        \text{For ATE: } (\hat{w_{1i}}, \hat{w_{0i}}) = \left(\frac{1}{\widehat{PS_i}}, \frac{1}{1 - \widehat{PS_i}}\right).
    \qquad
        \text{For ATT: } (\hat{w_{1i}}, \hat{w_{0i}}) = \left(1, \frac{\widehat{PS_i}}{1 - \widehat{PS_i}}\right).
    \end{equation*}   

    The IPW analysis have been conducted using the \textbf{PSweight} \textsf{R} package\cite{Zhou2022}.
    
    \item PSM. This approach aims to minimize selection bias by creating a balanced pseudo-population where treated and untreated individuals share similar PS\cite{Greifer2025, cenzer2020, ho2007}. While standard matching algorithms (e.g., nearest neighbor matching) typically achieve balance by discarding unmatched control units, a process that restricts inference to the ATT, we specifically selected full matching (PSM-FM) for this analysis. This method combines matching and stratification and allows the estimation of both the ATE and the ATT, optimally partitioning the entire sample into $K$ disjoint subclasses or matched sets ($S_1, \dots, S_K$) such that each set contains at least one treated ($n_{1k} \geq 1$) and one control ($n_{0k} \geq 1$) unit, ensuring no observations are discarded. By applying specific weights to these strata, PSM-FM can validly estimate both ATE and ATT.
    
    To calculate these weights, we can compute a stratum-specific PS as the empirical probability of treatment within each subclass:
    \begin{equation*} 
        \widehat{PS}_{S_k}^* = \hat{P}(A=1 \mid S = S_k) = \frac{n_{1k}}{n_k}, 
    \end{equation*}
    where $n_k$ is the total number of individuals in stratum $S_k$, and $n_{1k}$ is the count of treated individuals in that stratum. Consequently, the weights of the $i$-th individual within stratum $k$ depending on whether or not they received the treatment, $\hat{w}_{1k(i)}$ and $\hat{w}_{0k(i)}$, are calculated analogously to the IPW approach but using this stratum-specific probability.

    \begin{equation*}
        \text{For ATE: }(\hat{w}_{1k(i)}, \hat{w}_{0k(i)}) = \left(\frac{1}{\widehat{PS}_{S_k}^*}, \frac{1}{1-\widehat{PS}_{S_k}^*} \right). 
        \qquad
        \text{For ATT: } (\hat{w}_{1k(i)}, \hat{w}_{0k(i)}) = \left(1, \frac{\widehat{PS}_{S_k}^*}{1 - \widehat{PS}_{S_k}^*}\right).
    \end{equation*}

    This weighting scheme reconstructs the original population structure, distinguishing it from strategies that only target 
    the ATT\cite{Greifer2025, Austin2017}.   

    The expectation of the counterfactual worlds, $\hat{E}(Y^1)$ and $\hat{E}(Y^0)$, are computed like in the case of IPW, but with the specific weights calculated for PSM-FM. The PSM-FM procedure  have been executed using the \texttt{matchit} function within the \textbf{MatchIt} \textsf{R} package\cite{MathchIt} and the effects have been estimated using the \textbf{marginaleffects} \textsf{R} package\cite{Marginal_Effects}. 
   
    \item G-comp. It can be viewed as a model-based extension of outcome standardization. This method estimates the causal effect by explicitly modeling the conditional expectation of the outcome given the treatment and covariates $E(Y\mid A,L)$\cite{Greifer2025, R, Smith2022, Wang2017}. It relies on the consistency and exchangeability assumptions to impute the unobserved potential outcomes. The procedure involves fitting a regression model to the observed data and using it to predict the expected outcome for every individual under both treatment and control scenarios.
    
    To estimate the ATE, predictions are generated for the entire sample under each treatment scenario $a \in \{0, 1\}$. To estimate the ATT, the averaging of these predictions is restricted to the subpopulation of individuals who actually received the treatment. The expectations for each estimand are calculated as shown in Equations \ref{eq:gcomp1} and \ref{eq:gcomp2}, where $n_1$ is the number of treated individuals.
    \begin{equation}
    \label{eq:gcomp1}
            \text{For ATE: }\quad \hat{E}(Y^a) = \frac{1}{n} \sum_{i = 1}^n \hat{P}(Y_i^a = 1 \mid  L_i), \quad \text{for } a \in \{0, 1\}.
    \end{equation}
    \begin{equation}
    \label{eq:gcomp2}
           \text{For ATT: }\quad \hat{E}(Y^a\mid A=1) = \frac{1}{n_1} \sum_{i = 1}^n \hat{P}(Y_i^a = 1 \mid L_i) \cdot A_i, \quad \text{for } a \in \{0, 1\}.
    \end{equation}

    The causal effects  are subsequently obtained by comparing these standardized probabilities (e.g., on the risk difference scale, $\text{ATE} : \hat{E}(Y^1) - \hat{E}(Y^0)$). G-computation have been performed using the \texttt{avg$\_$comparisons} function of the \textbf{marginaleffects} \textsf{R} package\cite{Marginal_Effects}.

    \item TMLE. It is a semiparametric doubly robust method that combines the models for the treatment assignment $E(A\mid L) = P(A = 1 \mid L) = PS$ and the outcome mechanism $Q(A, L) = E(Y\mid A,L)$\cite{TMLE}. 
    
    The procedure involves two stages. First, initial estimates of the outcome, $Q^0(A,L)$, and the PS are calculated. Second, $H(0,L)$ and $H(1,L)$ are defined as clever covariates. These covariates are derived from the estimate $PS$ as shown in Equation \ref{eq:tmle1}.
    
    \begin{equation}
    \label{eq:tmle1}
        H(1,L) = \frac{A}{\widehat{PS}}, \qquad H(0,L) =  \frac{1-A}{1-\widehat{PS}}.
    \end{equation} 
    
    It should be noted that these clever covariates are mathematically equivalent to the weights used in the IPW framework.
    Then, a vector fluctuation parameter, denoted by $\varepsilon = (\varepsilon_0, \varepsilon_1)$, is estimated to update the initial outcome estimates $Q^0(A,L)$. $\varepsilon$ is estimated by maximum likelihood as the regression coefficients associated with the clever covariates in an intercept-free logistic regression for the outcome $Y$, using $\text{logit}(Q^0(A,L))$ as an offset. This \emph{targeting} step reduces bias by updating the initial outcome estimate using the fluctuation parameter $\varepsilon$. The PS is used to guide this adjustment toward the causal parameter of interest\cite{luque2018}. 
    
    The updated estimate of the parameter, denoted by $Q^1(A,L)$, is obtained by replacing  the fluctuation parameter into the Equations \ref{eq:tmle2} and \ref{eq:tmle3}, where $\operatorname{expit}\!\left(x\right) = 1 / {\left(1+e^{-x}\right)}$. 
    \begin{equation}
    \label{eq:tmle2}
        \hat{Q}^1(0, L)
        = \operatorname{expit}\!\left(
        \operatorname{logit}\bigl(\hat{Q}^0(0, L)\bigr)
        + \frac{\hat{\varepsilon}_0}{1 - \widehat{PS}}
        \right), 
    \end{equation}
    \begin{equation}
    \label{eq:tmle3}
        \hat{Q}^1(1, L)
        = \operatorname{expit}\!\left(
        \operatorname{logit}\bigl(\hat{Q}^0(1, L)\bigr)
        + \frac{\hat{\varepsilon}_1}{\widehat{PS}}
        \right),
    \end{equation}
     The final causal effect is obtained by comparing the targeted marginal predictions over the full sample (ATE) or over the treated subpopulation (ATT).
    \begin{align*}
            &\text{For ATE: }\quad \hat{E}(Y^a) = \frac{1}{n} \sum_{i = 1}^n \hat{Q}^1(a, L_i), \quad \text{for } a \in \{0, 1\}.\\
           &\text{For ATT: }\quad \hat{E}(Y^a\mid A=1) = \frac{1}{n_1} \sum_{i = 1}^n \hat{Q}^1(a, L_i) \cdot A_i, \quad \text{for } a \in \{0, 1\}.
    \end{align*}
    TMLE provides consistent estimates if at least one of the two mechanisms is correctly specified and is asymptotically efficient when both are correct\cite{TMLE_foundation,Smith2022}. This methodology have been implemented using the \texttt{TMLE} function from the \textsf{R} package \textbf{TMLE} \cite{TMLE}. However, a limitation of the current version of this package is that it does not return the estimation of the variance of the OR for the ATT. To address this, we extended the standard implementation by developing our own function. This personalized function builds upon the core algorithm of the \texttt{TMLE} function to explicitly compute and output the required ATT OR estimates. Given the estimation procedures involved, a seed have been set to ensure the reproducibility of the results.
\end{itemize}

Uncertainty associated with the causal estimators considered in this study can be computed using sandwich estimators via the delta method \cite{Greifer2025}. This approach typically yields wider and more conservative confidence intervals (CI) while allowing control for data aggregation. Alternatively, bootstrap procedures can be used; these generally produce narrower intervals but are computationally more intensive, may introduce issues related to data aggregation, and are not theoretically supported for TMLE. In this manuscript, we opted for the delta method with sandwich estimators due to its broader applicability and faster computational implementation. In Table \ref{tab:rd_syntax}, the reader can find the \textsf{R} \cite{R} syntax for implementing RD estimators for the ATE and the ATT \cite{luque2018, Smith2022, Marginal_Effects, Zhou2022, MathchIt}.

\subsection{Data generating mechanism}

\begin{figure}[H]
    \centering
    \includegraphics[width=0.85\textwidth]{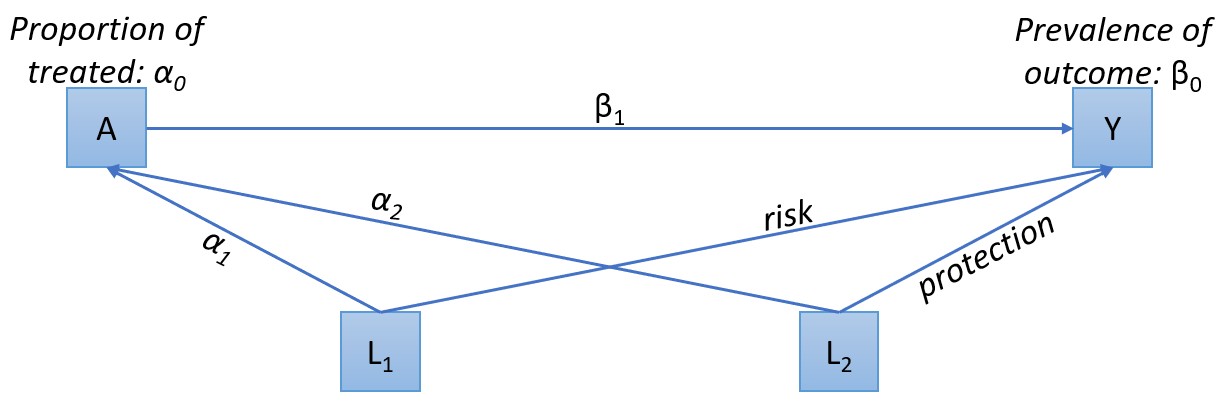}
    \caption{Directed acyclic graph representing the data generating mechanism.}
    \label{fig:simulacion}
\end{figure}

\begin{table}[H]
\caption{\textsf{R} syntax for obtaining RD and its SE of ATE and ATT. RD: risk difference. SE: Standard error.}
\centering
\scriptsize
\begin{tabularx}{\textwidth}{p{3.2cm} X X}
\toprule
\textbf{Method} & \textbf{ATE} & \textbf{ATT} \\
\midrule

\textbf{IPW} \\ \texttt{library(PSweight)} & 
\texttt{IPW\_est <- PSweight(} \newline
\hspace*{5pt}\texttt{ps.formula = A $\sim$ L1 + L2,} \newline
\hspace*{5pt}\texttt{data = as.data.frame(data),} \newline
\hspace*{5pt}\texttt{weight = "IPW",} \newline 
\hspace*{5pt}\texttt{yname = "Y",} \newline 
\hspace*{5pt}\texttt{trtgrp = "1")} \newline
\texttt{IPW\_ATE <- summary(IPW\_est)} \newline
\texttt{IPW\_ATE\$estimates[1] \# ATE} \newline
\texttt{IPW\_ATE\$estimates[2] \# ATE SE}
& \texttt{IPW\_est\_TT <- PSweight(} \newline
\hspace*{5pt}\texttt{ps.formula = A $\sim$ L1 + L2,} \newline
\hspace*{5pt}\texttt{data = as.data.frame(data),} \newline
\hspace*{5pt}\texttt{weight = "treated",} \newline
\hspace*{5pt}\texttt{yname = "Y",} \newline 
\hspace*{5pt}\texttt{trtgrp = "1")} \newline
\texttt{IPW\_ATT <- summary(IPW\_est\_TT)} \newline
\texttt{IPW\_ATT\$estimates[1] \# ATT} \newline
\texttt{IPW\_ATT\$estimates[2] \# ATT SE} \\

\midrule

\textbf{PSM-FM} \\ 
\texttt{library(marginaleffects)} \newline 
\texttt{library(MatchIt)} & 
\texttt{m.out <- matchit(
\newline
\hspace*{5pt}\texttt{A $\sim$ L1 + L2,} 
\newline 
\hspace*{5pt}\texttt{method = "full",}
\newline
\hspace*{5pt}\texttt{estimand = "ATE",}} 
\newline
\hspace*{5pt}\texttt{data = data)}
\newline
\texttt{mdata\_ATE <- match.data(m.out)} \newline
\texttt{log\_ATE <- glm(}
\newline
\hspace*{5pt}\texttt{Y $\sim$ A,} 
\newline
\hspace*{5pt}\texttt{family = quasibinomial,}
\newline
\hspace*{5pt}\texttt{data = mdata\_ATE,} 
\newline
\hspace*{5pt}\texttt{weights = mdata\_ATE\$weights)} 
\newline
\texttt{PSM\_ATE <- avg\_comparisons(}
\newline
\hspace*{5pt}\texttt{log\_ATE, }
\newline
\hspace*{5pt}\texttt{variables = "A", }
\newline
\hspace*{5pt}\texttt{vcov =~subclass)} 
\newline
\texttt{PSM\_ATE\$estimate \# ATE} \newline
\texttt{PSM\_ATE\$std.error \# ATE SE}
& \texttt{m.out.TT <- matchit(}
\newline
\hspace*{5pt}\texttt{A $\sim$ L1 + L2, }
\newline
\hspace*{5pt}\texttt{method = "full",} 
\newline
\hspace*{5pt}\texttt{estimand = "ATT",} 
\newline
\hspace*{5pt}\texttt{data = data)} 
\newline
\texttt{mdata\_ATT <- match.data(m.out.TT)} 
\newline
\texttt{log\_ATT <- glm(}
\newline
\hspace*{5pt}\texttt{Y $\sim$ A,} 
\newline
\hspace*{5pt}\texttt{family = quasibinomial, }
\newline
\hspace*{5pt}\texttt{data = mdata\_ATT,} 
\newline
\hspace*{5pt}\texttt{weights = mdata\_ATT\$weights)} 
\newline
\texttt{PSM\_ATT <- avg\_comparisons(}
\newline
\hspace*{5pt}\texttt{log\_ATT, }
\newline
\hspace*{5pt}\texttt{variables = "A",} 
\newline
\hspace*{5pt}\texttt{vcov =~subclass, }
\newline
\hspace*{5pt}\texttt{newdata = subset(A == 1))}
\newline
\texttt{PSM\_ATT\$estimate \# ATT} \newline
\texttt{PSM\_ATT\$std.error \# ATT SE} \\

\midrule

\textbf{G-computation} \\ 
\texttt{library(marginaleffects)} & 
\texttt{fit1 <- glm(}
\newline 
\hspace*{5pt}\texttt{Y $\sim$ A * (L1 + L2), }
\newline 
\hspace*{5pt}\texttt{family = "binomial",} 
\newline 
\hspace*{5pt}\texttt{data = data)} 
\newline
\texttt{G\_comp\_ATE <- avg\_comparisons(}
\newline
\hspace*{5pt}\texttt{fit1, }
\newline
\hspace*{5pt}\texttt{variables = "A")} 
\newline
\texttt{G\_comp\_ATE\$estimate \# ATE} \newline
\texttt{G\_comp\_ATE\$std.error \# ATE SE}
& \texttt{G\_comp\_ATT <- avg\_comparisons(}
\newline
\hspace*{5pt}\texttt{fit1, }
\newline
\hspace*{5pt}\texttt{variables = "A",} \newline
\hspace*{5pt}\texttt{newdata = subset(A == 1))} 
\newline
\texttt{G\_comp\_ATT\$estimate \# ATT} \newline
\texttt{G\_comp\_ATT\$std.error \# ATT SE} \\

\midrule

\textbf{TMLE} \\ \texttt{library(tmle)}
& \texttt{set.seed(1) \# for reproducibility} 
\newline
\texttt{tmle\_AT <- tmle(}
\newline
\hspace*{5pt}\texttt{Y = data\$Y, }
\newline
\hspace*{5pt}\texttt{A = data\$A,} 
\newline
\hspace*{5pt}\texttt{W = data[, c("L1","L2")],}
\newline
\hspace*{5pt}\texttt{family = "binomial",} 
\newline
\hspace*{5pt}\texttt{Q.SL.library = "SL.glm", }
\newline
\hspace*{5pt}\texttt{g.SL.library = "SL.glm")} 
\newline
\texttt{tmle\_AT\$estimates\$ATE\$psi \# ATE} 
\newline
\texttt{sqrt(tmle\_AT\$estimates\$ATE\$var.psi) \# ATE SE}
& \texttt{tmle\_AT\$estimates\$ATT\$psi \# ATT} 
\newline
\texttt{sqrt(tmle\_AT\$estimates\$ATT\$var.psi) \# ATT SE} \\

\bottomrule
\end{tabularx}
\label{tab:rd_syntax}
\end{table}

We consider a data generating mechanism with a binary treatment $A$, a binary outcome $Y$, and two confounding covariates, $L_1$ and $L_2$. The assumed causal structure is represented by the directed acyclic graph shown in Figure \ref{fig:simulacion} \cite{Pearl1995}. 
The covariates $L_1$ and $L_2$ can be interpreted as linear combinations of baseline characteristics, where $L_1$ groups risk factors and $L_2$ represents protective factors for the outcome \cite{Austin2023}. Both covariates are independently generated from a normal distribution $N(\mu = 1, \sigma = 0.2)$, setting both parameters for simplicity. The probability of prescribing the treatment $A$ in a patient $i$ is determined by an inverse logit function,
$${PS}_i=\operatorname{expit}\!\left(\alpha_{0}+\alpha_{1}\cdot L_{1i}+\alpha_{2}\cdot L_{2i}\right),$$  
and the treatment assignment of an individual was defined by a Bernoulli distribution, $A_i \stackrel{\text{ind.}}{\sim} Bernoulli\!\left({PS}_i\right)$. 
The outcome model is designed to obtain a heterogeneous treatment effect. For this outcome model, the coefficients for $L_1$ and $L_2$ were set to $0.5$ and $-0.3$, respectively, to induce a reasonable level of variability in the outcome and allow fair comparisons with the magnitude of the treatment effect. 
The probability of the outcome in a counterfactual world where every patient is untreated, $P(Y^0 = 1)$, is defined by an inverse logit function,
$$P\left(Y_i^0=1\right)=\operatorname{expit}\!\left(\beta_{0}+0\cdot\beta_{1}+0.5\cdot L_{1i}-0.3\cdot L_{2i}\right),$$ 
where $\beta_1$ is the conditional log(OR) of the treatment. To introduce treatment effect heterogeneity, a necessary condition to ensure that the ATE and the ATT represent distinct target values, we defined the potential outcome under treatment, $P(Y^1 = 1)$, conditionally on the PS. Specifically, for individuals with a PS $< 0.5$, we assumed a null treatment effect (setting $\beta_1 = 0$); for these individuals, the risk under treatment is identical to the risk under control. 
Conversely, for individuals with a PS $\geq0.5$, the treatment has a non-zero effect ($\beta_1 \neq 0$), with the probability of the outcome defined by the inverse logit function,  $$P\left(Y_i^1=1\right)=\operatorname{expit}\!\left(\beta_{0}+\beta_{1}+0.5\cdot L_{1i}-0.3\cdot L_{2i}\right).$$ 

With this specification, we induce treatment effect heterogeneity. As a consequence, the ATE and the ATT will not coincide because the treatment effect varies across individuals and the distribution of covariates differs between the treated and untreated groups\cite{Pirracchio2016}.

The potential outcome in each counterfactual world is defined as $$Y_i^A \stackrel{\text{ind.}}{\sim} \text{Bernoulli}\!\left(P\left(Y_i^A = 1\right)\right).$$ By the consistency assumption, the observed outcome of a patient $i$ equals the potential outcome under the treatment that is actually received\cite{Whatif,luque2018}, $$Y_i=\ Y_i^1\cdot A_i+\ Y_i^0\cdot\left({1-A}_i\right).$$

\subsection{Design of experiments: Scenarios}

\paragraph{Positivity assumption violation}

To illustrate the positivity assumption violation problem, we have chosen different settings with different PS-OV between treated and untreated individuals. The smaller the PS-OV, the greater the degree of positivity assumption violation \cite{Whatif,Zhou2020}. The overlapping coefficient represents the total shared area between the two density functions. It is computed by integrating, at each point, the minimum of both densities:
$$\text{PS-OV} = 100\% \cdot \int_0^1 \min (f_{PS^0}(p), f_{PS^1}(p)) \, dp,$$
where $f_{PS^a}\left(x\right)$ is the PS density function conditional on the treatment group $A = a$ and $p$ denotes the values within the support of the PS, specifically the interval $[0, 1]$. The PS-OV ranges between 0\% (curves perfectly separated, i.e. total lack of common support) and 100\% (both curves are indistinguishable, i.e. identical distributions between groups)\cite{Alba2021, Franklin2014}. The coefficients $\alpha_1$ and $\alpha_2$ in the treatment selection model were determined numerically to obtain the desired PS-OV in the superpopulation, and the intercept $\alpha_0$ was also determined numerically to obtain the desired proportion of treated patients. 

To isolate violations of the positivity assumption arising from low PS-OV between treatment groups, the model specification was kept correct for each statistical technique. Additionally, PS and outcome were consistently estimated using logistic regression across all methods. In this context, a correctly specified model refers to a situation in which the set of included covariates correspond to the true data-generating mechanism, so that no bias is introduced due to model misspecification. In contrast, a misspecified model arises when relevant confounders are omitted, inappropriate functional forms are assumed, or incorrect link functions are used, potentially leading to biased causal effect estimates\cite{Whatif, luque2018}.

\begin{figure}[H]
    \centering

    \begin{minipage}[b]{0.32\textwidth}
        \centering
        \includegraphics[width=\textwidth]{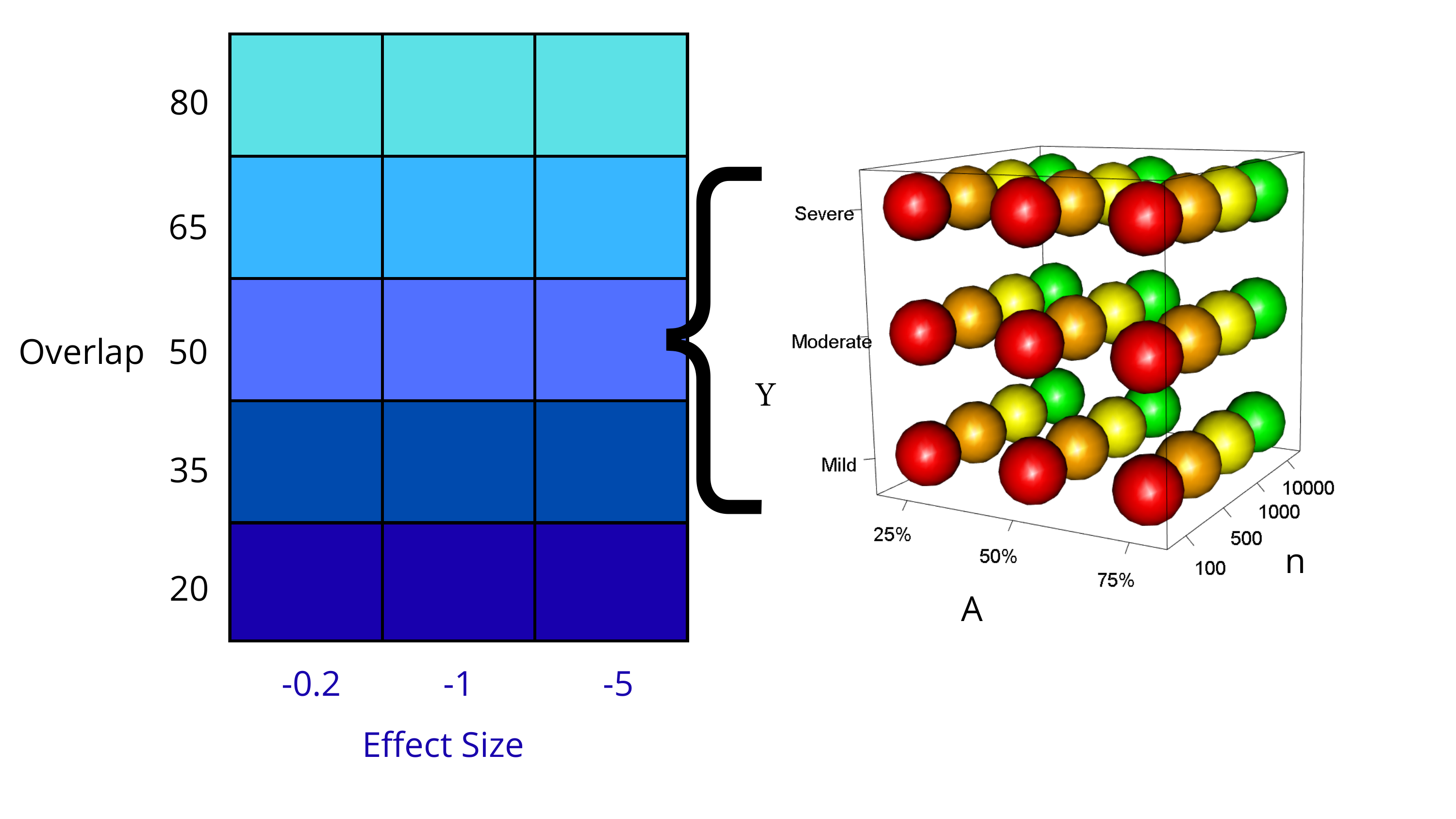}
    \end{minipage}
    \hfill
    \begin{minipage}[b]{0.32\textwidth}
        \centering
        \includegraphics[width=\textwidth]{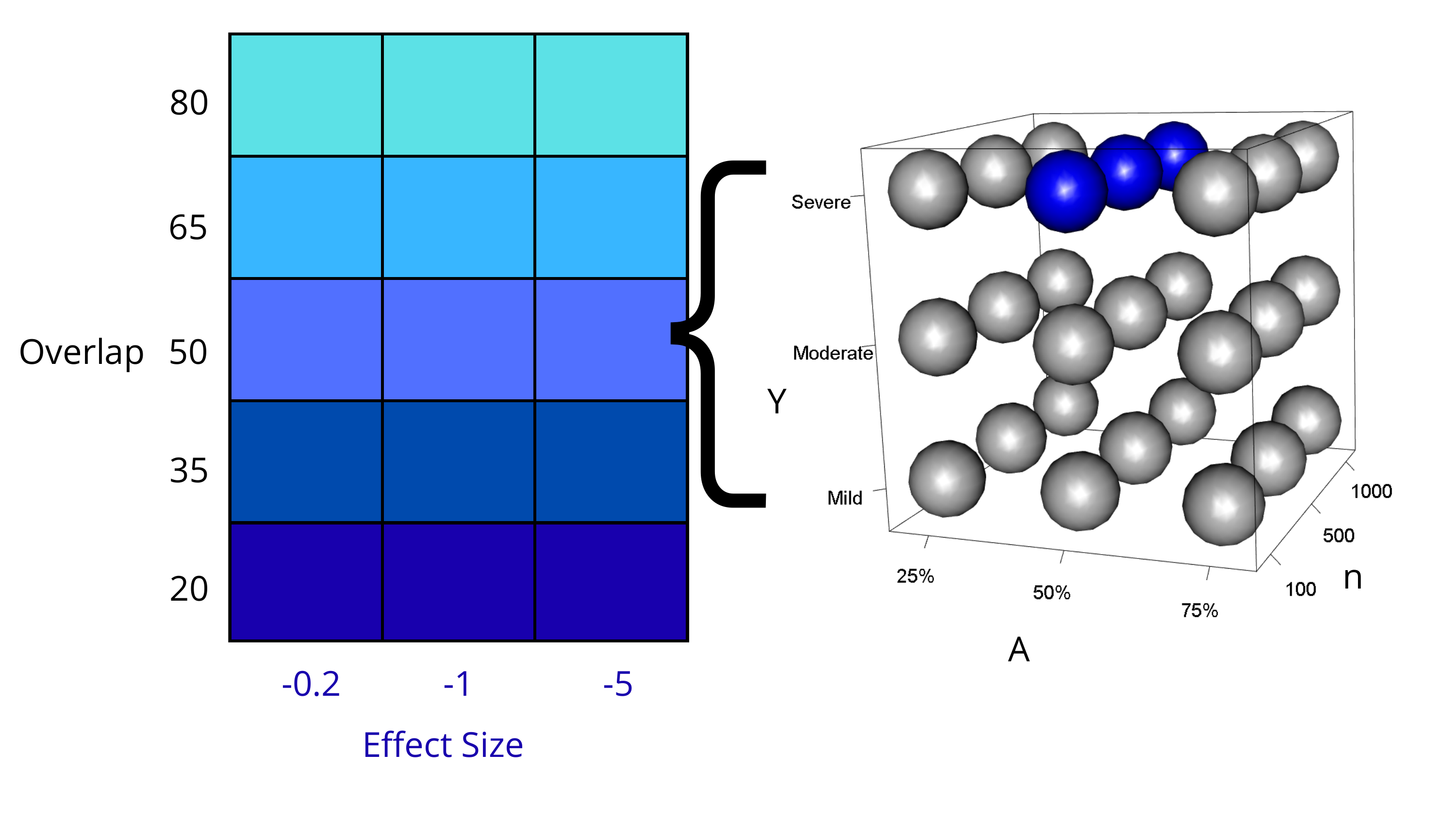}
    \end{minipage}
    \hfill
    \begin{minipage}[b]{0.32\textwidth}
        \centering
        \includegraphics[width=\textwidth]{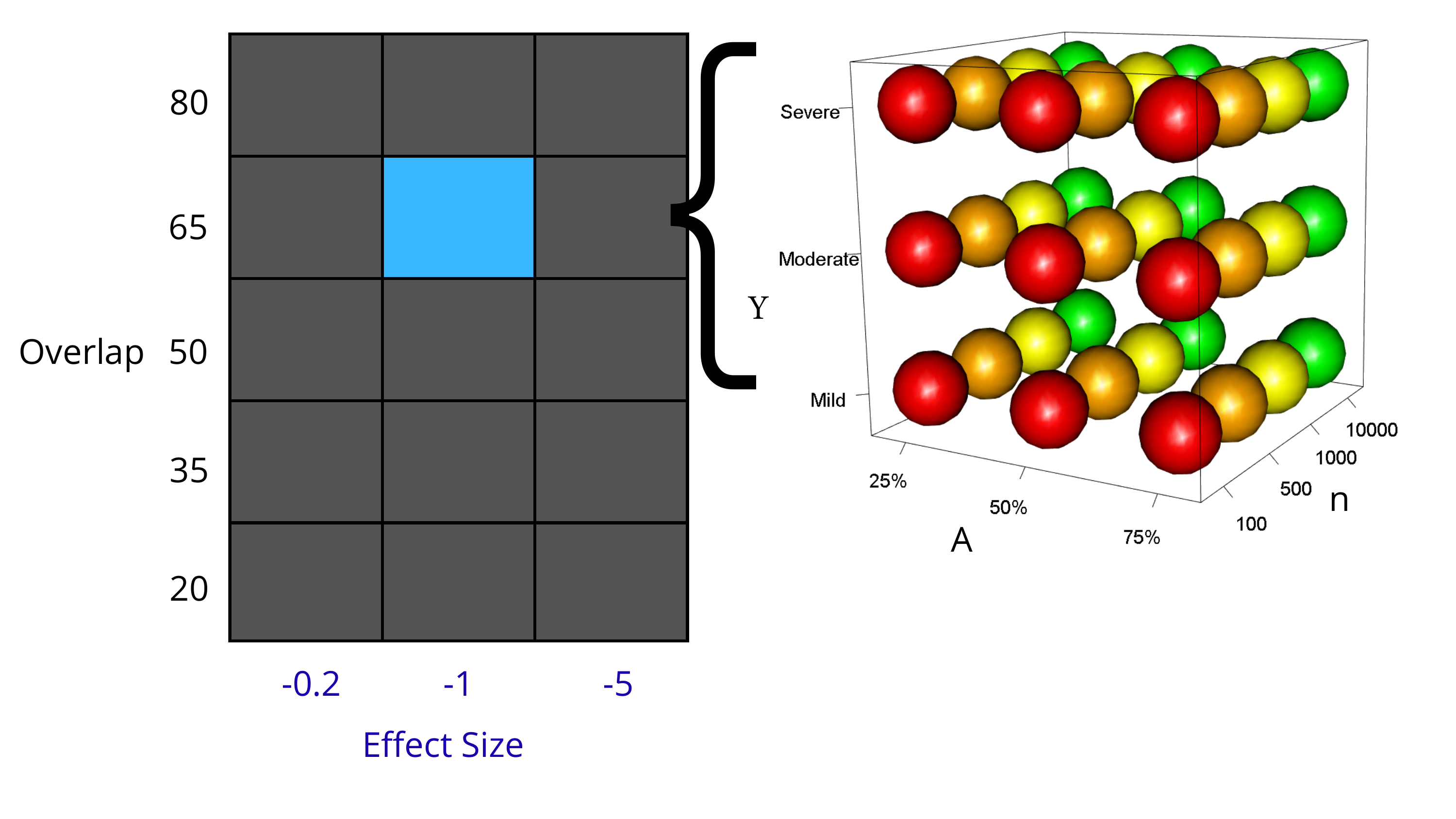}
    \end{minipage}
    \caption{Simulation study design. Left: Full set of simulations we aim to cover. Center: First stage of simulations. Right: Second stage of simulations. }
    \label{fig:schema_simulations}
\end{figure}

\paragraph{Scenarios considered in the design}
For the studied settings, this work considers all possible combinations of the parameter levels specified in Table \ref{tab:simulation_params}, where the selected parameter values are displayed. Causal effects were studied in absolute scale (RD) and relative scale (marginal OR). Figure~\ref{fig:schema_simulations} shows the design of the simulations.
An extensive description of the experiment design is detailed in Supplementary Section S.1. A wide range of scenarios has been developed, covering a variety of situations with respect to disease prevalence, treatment efficacy, and treatment allocation policies. Variations in allocation policies involve the fraction of the treated population and the political decision with respect to protecting high-risk individuals more intensely; note that this implies different PS-OV situations.

In the first stage, we have studied how positivity assumption violation affects to each chosen statistical method in a scenario with a fixed proportion of treated patients equal to 0.5 ($\alpha_0 = 0$), and a fixed proportion of the outcome also equal to 0.5 ($\beta_0 = 0$). The scenarios compared three different values for the effect size $\beta_1$, $-0.2$, $-1$, and $-5$, and five degrees of PS-OV, approximately 20, 35, 50, 65, and 80\%. Also, three different sample sizes were examined, $n =$ 100, 500, and 1000 individuals. In total, $3 \text{ (effect size)} \times 5 \text{ (PS-OV)} \times 3 \text{ (sample size)} = 45$ scenarios were examined for both ATE and ATT. 

A preliminary analysis of the first stage showed that high PS-OV allowed all methods to perform properly, whereas low PS-OV caused almost all methods to perform poorly, as expected under violations of the positivity assumption\cite{Whatif, Li2018}. Due to the consistently poor performance in low-overlap scenarios and the generally good performance in high-overlap scenarios, we decided not to further develop comparisons in these extreme cases, and instead focus on intermediate values of PS-OV, around 60\%, which showed different behavior among the different tested statistical techniques. 

Consequently, in the second stage, we selected a scenario with a fixed degree of PS-OV of 60\% and a moderate fixed effect size to investigate the impact of the distributional characteristics of treated patients and the outcome proportion. We have concentrated on the interplay between the proportions of treated patients: approximately 25\% ($\alpha_0 = -1.25$), 50\% ($\alpha_0 = 0$), and 75\% ($\alpha_0 = 1.25$); outcome prevalence, as different levels of disease severity or event frequency: mild (probability approximately 3 to 7\%, $\beta_0 = -2.5$), moderate (probability approximately 10 to 20\%, $\beta_0 = -1.5$), and severe (probability approximately 20 to 40\%, $\beta_0 = -0.6$); and sample size: small ($n = 100$), medium ($n = 500$), large ($n = 1000$), and very large ($n = 10{,}000$). Thus, $3 \text{ (proportions of treated patients)} \times 3 \text{ (outcome prevalence)} \times 4 \text{ (sample size)} = 36$ settings were examined both for ATE and ATT. 

Table~\ref{tab:simulation_params} shows a summary of the selected parameters in both stages.

\begin{table}[ht]
    \centering
    \caption{Simulation scenarios and coefficient values for the first and second stages of the study. In the first stage, sample sizes of $n = 100, 500 \text{ and } 1000$ are considered, while the second stage also includes $n = 10,000$.}
    \label{tab:simulation_params}
    \begin{minipage}{0.48\textwidth}
        \centering
        \begin{tabular}{ccc}
            \toprule
            \multicolumn{3}{c}{\textbf{First stage}, $\alpha_0 = 0$, $\beta_0 = 0$} \\ \midrule
            \textbf{PS-OV} & $\alpha_1 = \alpha_2$ & $\beta_1$ \\ \midrule
            80 & 1.85  & $-0.2, -1, -5$ \\
            65 & 3.54  & $-0.2, -1, -5$ \\
            50 & 5.90  & $-0.2, -1, -5$ \\
            35 & 10.60 & $-0.2, -1, -5$ \\
            20 & 22.61 & $-0.2, -1, -5$ \\ \bottomrule
        \end{tabular}
    \end{minipage}
    \hfill 
    \begin{minipage}{0.48\textwidth}
        \centering
        \begin{tabular}{ccc}
            \toprule
            \multicolumn{3}{c}{\textbf{Second stage}, $\alpha_1 = \alpha_2 = 4.25$, $\beta_1 = -1$} \\ \midrule
            \textbf{PS-OV} & $\alpha_0$ & $\beta_0$ \\ \midrule
            60 & $-1.25$ & $-2.5, -1.5, -0.6$ \\
            \\
            60 & 0     & $-2.5, -1.5, -0.6$ \\
            \\
            60 & 1.25  & $-2.5, -1.5, -0.6$ \\ \bottomrule
        \end{tabular}
    \end{minipage}
\end{table}

\paragraph{Protective effect of the treatment}

For the outcome model, the coefficient $\beta_1$ was chosen to obtain different strengths of the protective treatment effect, according to its conditional log(OR), as mentioned above. In Supplementary Section S.2, Figures S.4 and  S.5 the reader can see the heterogeneous treatment effect comparing both counterfactual worlds. For each square in the heatmap, the probability of the outcome if the individual is treated is compared with the probability of the outcome if the same individual is not treated (real RD). This is based on the individual's baseline characteristics (defined as the standard deviations of $L_1$ and $L_2$) and other characteristics shown in the heat map. As the data generating mechanism is known, we can compute exactly this effect. As an example, for the first stage of the simulation, an individual who receive a treatment with the strongest protective effect, has a risk factor ($L_1$) one standard deviation (SD) above the mean, and a protective factor ($L_2$) two SD below the mean, has an absolute reduction of the probability of outcome equal to 0.593, compared with the same untreated individual in a counterfactual world. Real RD for each individual does not depend on PS-OV and can rarely be calculated in real world settings, except for crossover experiments under very strong additional assumptions\cite{Whatif}.

\paragraph{Superpopulation simulation for true causal effects}

To determine the true PS-OV between groups, the proportion of treated patients, the proportion of patients with the outcome, and the true marginal causal estimates (ATE and ATT), these quantities were computed in an arbitrarily large superpopulation ($n = 10^7$). The syntax used is provided in Supplementary Section S.2, Code box.

Because counterfactual outcomes are known, the true ATE and ATT were calculated for the RD as:
\begin{align*}
ATE_{RD} &= E(Y^1) - E(Y^0),\\
ATT_{RD} &= E(Y^1 \mid A = 1) - E(Y^0 \mid A = 1);
\end{align*}
and for the marginal OR\cite{luque2018,morris2019,Wang2025,Zhou2022}:
\begin{align*}
ATE_{OR} &= \frac{E(Y^1)/ (1-E(Y^1))}{E(Y^0) / (1-E(Y^0))},\\
ATT_{OR} &= \frac{E(Y^1 \mid A=1)/ (1-E(Y^1 \mid A=1))}{E(Y^0 \mid A=1) / (1-E(Y^0 \mid A=1))}.
\end{align*}

\paragraph{Monte Carlo simulations}

We simulated samples in every scenario to perform a Monte Carlo simulation. To determine the required number of repetitions (\texttt{nsim}) of the Monte Carlo simulation for each stage, we performed first a preliminary analysis with 100 samples to obtain empirical standard error in each stage (\textit{EmpSE}) and computed its Monte Carlo standard error (MCSE). Then, we computed the desired number of \texttt{nsim} to achieve an acceptable MCSE. The number of simulated samples required to obtain a Monte Carlo standard error for the estimated bias below 0.001 was 1067 for the first stage, and 610 for the second stage\cite{morris2019}. 

\paragraph{Summaries of simulation results}

We tabulate the performance measures with tables and figures comparing distinct scenarios for ATE and ATT. Although we calculated both RD and marginal OR for both stages, we chose to communicate RD for the first stage and marginal OR for the second stage. This decision was made to prevent over-information and ensure the clarity of the presentation, given the large volume of results generated across the multiple simulation scenarios. All analyses were conducted in \textsf{R} version 4.5.2\cite{R}, in accordance with the guidelines for fair simulation studies proposed by Morris et al. \cite{morris2019}. The main performance measure was bias. Other studied performance measures were relative bias, mean squared error (MSE) and coverage, with the construction of zipper plots. The point estimates and MCSEs were calculated for all performance measures. If the methods did not converge, the proportion of cases in which this occurred was followed by its 95\% CI, calculated using the Clopper-Pearson method. All the syntax in \textsf{R} can be found in GitHub repository https://github.com/aaurensanz/code-causal-inference-comparison.

\section{Results}\label{sec:3}

\subsection{First stage: wide range of scenarios}

Almost all methods produced results for all tested settings. Out of the 360 procedures examined (45 scenarios, 4 statistical techniques and 2 target estimands, ATE and ATT), only thirteen showed convergence issues. All methods gave results when computing RD for ATE and ATT. However, when computing marginal OR for ATE and ATT, IPW and TMLE sometimes failed to produce results, as can be seen in Supplementary Table S.5. This situation occurred with a very low frequency and only for IPW and TMLE, ranging generally 0.09\%  to 2.91\%. It occurred more frequently when the PS-OV was lower and the sample size was smaller. Almost all non-convergent events occur with the highest magnitude of the treatment effect. Only for IPW and a very extreme scenario (sample size 100, PS-OV = 20\%, and conditional log(OR) $= -5$), this probability was higher, reaching 7.03\%.

For ATE, bias was low for PS-OV over 50\% but increases quickly for lower PS-OV. G-comp showed an increasing bias earlier with decreasing PS-OV when the effect of the treatment was stronger, for PS-OV equal or lower than 65\%; and a similar behavior was observed with relative bias. Although PSM-FM showed a lower bias for low PS-OV, its MSE was the highest with PS-OV equal to 35\% and 20\%. Regarding coverage, over-coverage was not an issue for any statistical technique. All methods performed well with a high PS-OV, except for G-comp and the strongest treatment effect, where an under-coverage began to appear for PS-OV equal to 65\%. This situation is worse with the highest sample size. For low PS-OV, all methods had a deep under-coverage, that was less notorious for IPW and weak to moderate treatment effect. Moreover, for G-comp, coverage was poor for the moderate and strongest treatment effect for PS-OV equal or under 50\%. G-comp performed better for the weakest treatment effect. The other methods also had a bad performance, with also risk of biased estimations with growing treatment effect and sample size. Results for marginal OR were similar. All these results are presented for a deeper examination in Supplementary Section S.3, Figures S.6 to S.19. 

For ATT, the behavior of the tested statistical techniques was different. Bias increased with a slightly lower PS-OV (equal or below 35\%), G-comp registered the lowest bias and TMLE appears to be less robust in these settings. Relative bias showed similar results. MSE was low for PS over 50\%, but increased dramatically for IPW and, above all, for PSM-FM and TMLE, for a PS-OV under 50\%. Over-coverage was not either an issue for any method. The coverage was good for PS-OV over 50\%, except for small sample sizes and weak treatment effect for TMLE and PSM-FM. The coverage deteriorates quickly under this value, apart from G-comp, that achieves a good coverage for sample size over 500, regardless of PS-OV. Another major difference compared to ATE is that estimations from G-comp were not biased, but they are biased sometimes for TMLE and PSM-FM with low values of PS-OV. Results for marginal OR were also similar. All these results can be examined in detail in Supplementary Figures S.20 to S.33.

Thus, the behavior of each method was different depending on target population (ATE vs ATT), strength of the treatment effect, sample size, and degree of positivity assumption violation. For ATE, all methods performed poorly for low PS-OV (equal or below 50\%), and G-computation also gave sometimes biased estimations. TMLE and IPW were the most robust methods in those settings. For ATT, G-comp resisted more against low PS-OV than the rest of the statistical techniques. To summarize, general recommendations can be seen in Figure \ref{first_recomendation} and heat map displayed in Supplementary Figure S.34. Generally, a high PS-OV leads to good performance of all statistical techniques, whereas a low PS-OV commonly leads to poor performance. For a medium PS-OV, performance is different between statistical techniques and varies with characteristics of the sample. So, it is worth to develop a deeper investigation in these scenarios with an intermediate PS-OV.

\afterpage{ 
\begin{landscape}
  \begin{figure}[p] 
    \centering
    \includegraphics[height=0.42\textheight, width=\linewidth, keepaspectratio]{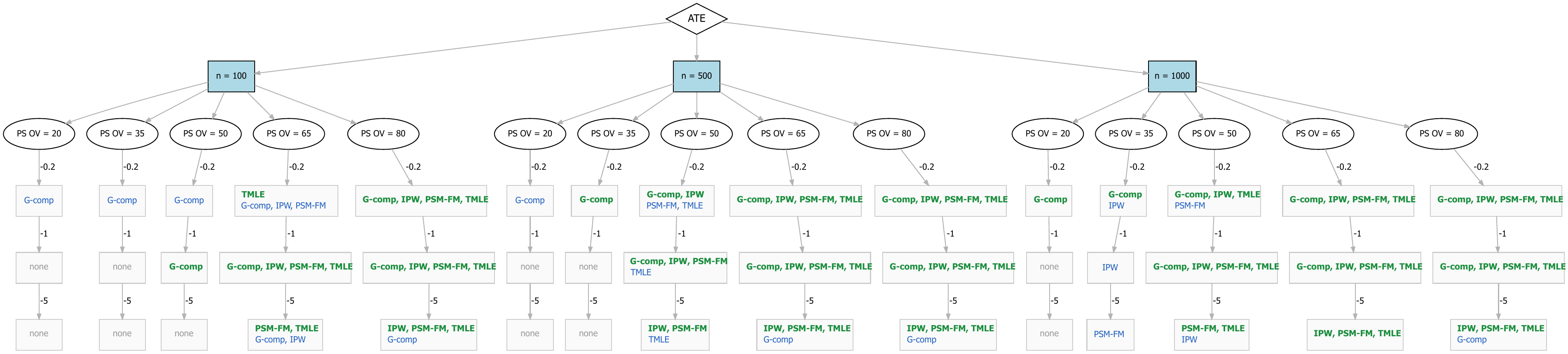}
    \vfill 
    \includegraphics[height=0.42\textheight, width=\linewidth, keepaspectratio]{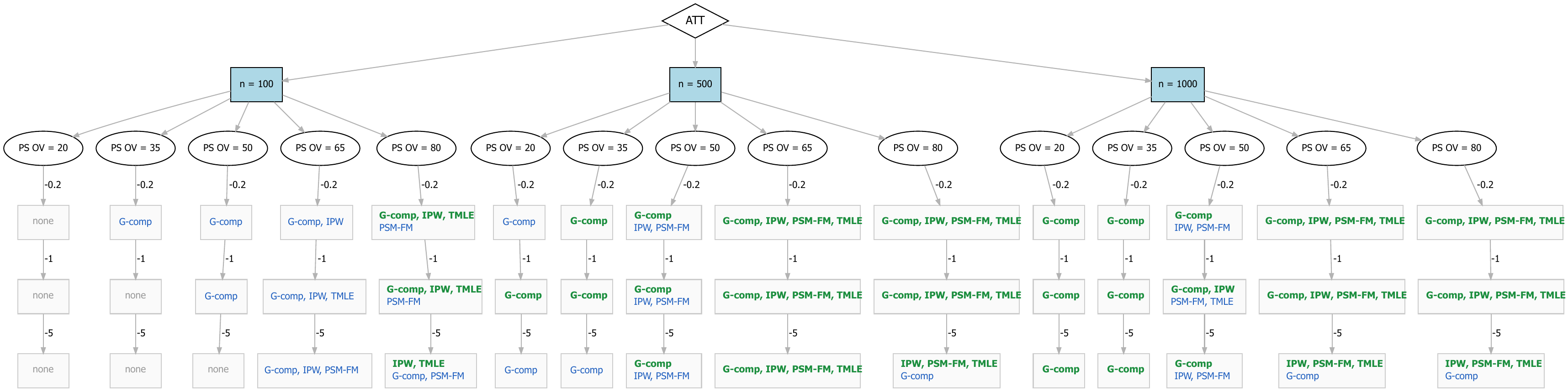}
    \caption{Suggested recommendations from the first stage of the simulation. Green indicates the best-performing methods, while blue denotes acceptable but lower-performing alternatives.} \label{first_recomendation}
  \end{figure}
\end{landscape}
}

\newpage

\subsection{Second stage: a common setting with PS-OV around 60\%}

Based on the initial assessment where severe positivity violations were shown to compromise the reliability of causal estimators, this section focuses on scenarios representative of well designed observational research characterized by moderate PS-OV (around 60\%). This approach seeks to focus on realistic practice scenarios in applied observational research, where an appropriate study design would prevent researchers from reaching low PS-OV values that lead to nearly violation of positivity assumption.  The true causal effect is fixed at a moderate protective level, while the proportion of treated individuals, outcome prevalence, and sample size are varied. The displayed results, as mentioned previously, are on a relative scale (marginal OR).

Due to the limitations of the smallest sample size ($n = 100$), a subset of iterations resulted in \textit{zero-event scenarios}. In those settings, none of the tested statistical techniques produced valid results, representing a specific challenge for the stability of the estimators. Some of them are characterized by a total absence of outcome events ($\sum Y_i = 0$) and were only observed in settings with low outcome prevalence (mild disease) and high treatment percentage (75\%), occurring in a percentage of 0.9\% (95\% CI: 0.4\% to 2.1\%). Beyond scenarios with a complete absence of events in the entire dataset, we identified iterations where no events were observed in either the treated group ($\sum_{A_i=1} Y_i = 0$) or the control group ($\sum_{A_i=0} Y_i = 0$). These scenarios appeared mainly in mild‑disease settings, with percentages varying by treatment prevalence between 8.5\% and 15.7\%. In moderate‑disease scenarios, their occurrence was uncommon, with percentages $\leq$0.7\%, as shown in the Supplementary Section S.4, Table S.6. Subsequent performance analysis excluded these iterations associated with \textit{zero-event scenarios} where causal estimation was computationally infeasible due to a lack of events. For the rest of studied settings, all methods always converged and gave valid results.

For the ATE, with respect to relative bias, most methods exhibited stable behavior in large samples ($n$ from 1000 to 10,000), with relative biases close to zero, whereas G-comp consistently showed elevated relative bias. As the sample size decreased, IPW and TMLE emerged as the most robust methods for minimizing relative bias when the proportion of treated individuals was between 25 and 50\%. In these settings, PSM-FM and G-comp exhibited consistently higher biases, making them less optimal choices. In contrast, under high treatment prevalence (75\%), G-comp achieved the lowest relative bias, an effect that became more pronounced as the sample size decreased.

In terms of MSE, all methods showed robust, comparable performance in very large samples; however, as sample size decreased, PSM-FM consistently yielded high MSE and therefore should be avoided in scenarios with moderate or small sample sizes. For moderate to large samples ($n$ from approximately 500 to 1000), IPW and TMLE achieved the lowest MSE under low treatment prevalence (25\%), whereas G-comp outperformed others in balanced or high prevalence of treatment scenarios. In small sample scenarios, IPW, G-comp and TMLE perform comparably with respect to MSE, and only PSM-FM obtains higher MSE values and should therefore be avoided.

Focusing on coverage, in very large samples with low treatment prevalence (25\%), G-comp demonstrated poor coverage due to bias and low variance, whereas other methods remained valid. Conversely, in balanced or high prevalence settings, all methodologies achieved satisfactory nominal coverage. At moderate to large samples, all methodologies maintained acceptable coverage, although IPW and TMLE proved most robust, consistently yielding probabilities closest to the nominal 95\% level. With small sample sizes, coverage fell below nominal levels for all methodologies, reflecting the inherent limitations of sparse data. All the results can be seen in Supplementary Figures S.35 to S.49.

For ATT, given the similarities in the structural behavior of the estimators, and to avoid redundancy, we summarize the results by focusing on the key deviations from the ATE findings. In very large sample scenarios, all four methodologies demonstrated robust estimation properties. Notably, G-comp, which exhibited substantial bias for the ATE, performed satisfactorily for the ATT, as was observed in the first stage of the simulation.

In moderate to large sample sizes, the comparative ranking of the methods shifted. While IPW and TMLE remained robust and valid options, G-comp emerged as the superior strategy, consistently achieving lower MSE and relative bias than the weighting-based and doubly robust approaches, as we also found in the first stage of the simulation.

Finally, in small sample settings, the optimal methodology becomes highly sensitive to the event rate and treatment prevalence. In scenarios with higher treatment prevalence (from 50 to 75\%) combined with a low outcome prevalence (mild disease/rare events), TMLE is the recommended methodology. In these specific conditions, G-comp suffered from model instability due to the scarcity of events. However, for all other small-sample scenarios (e.g., balanced or high prevalence of treatment designs, or moderate-to-high outcome prevalence), G-comp proved to be the most reliable estimator, offering the best stability relative to the available information. All the results can be seen in Supplementary Figures S.50 to S.63. 
The results obtained for the RD were highly consistent with those observed for the marginal OR across most evaluated settings. 

Notable exceptions were found in two specific scenarios: concerning ATE, when the sample size was small ($n = 100$) combined with a high treatment prevalence (75\%), no single methodology consistently outperformed the others. Regarding ATT, G-comp emerged as the most robust methodology, generally yielding the superior performance, particularly when evaluated through the MSE.

Based on previous results, we can derive the evidence-based handbook for estimating ATE and ATT in observational studies with moderate PS-OV, represented in Figure \ref{second_recomendation} and heat map displayed in Supplementary Figure S.64. Overall, for the ATE, IPW and TMLE consistently show the best performance across most scenarios. In contrast, for the ATT, G-comp exhibits a better overall performance, outperforming the alternative methods in a wide range of settings. For both ATE and ATT, the proportion of treated patients and the prevalence of the outcome were important for the performance of statistical techniques only for small samples ($n = 100$). For larger sample sizes, neither were important.

\afterpage{ 
\begin{landscape}
  \begin{figure}[p] 
    \centering
    \includegraphics[height=0.42\textheight, width=\linewidth, keepaspectratio]{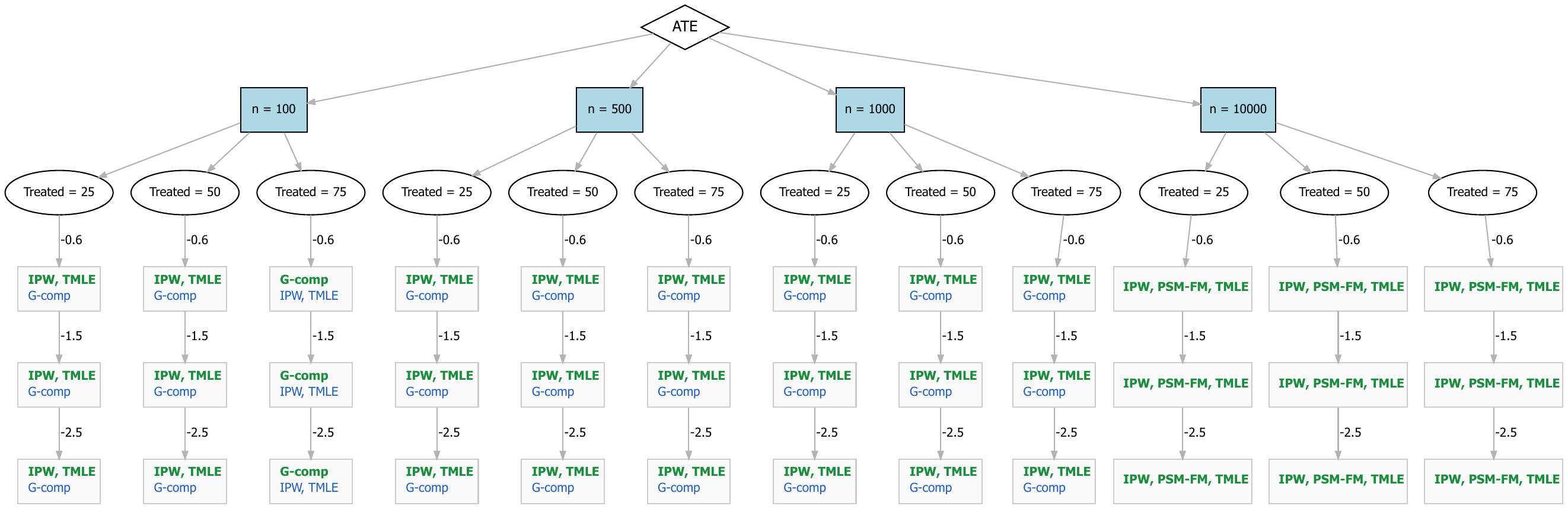}
    \vfill 
    \includegraphics[height=0.42\textheight, width=\linewidth, keepaspectratio]{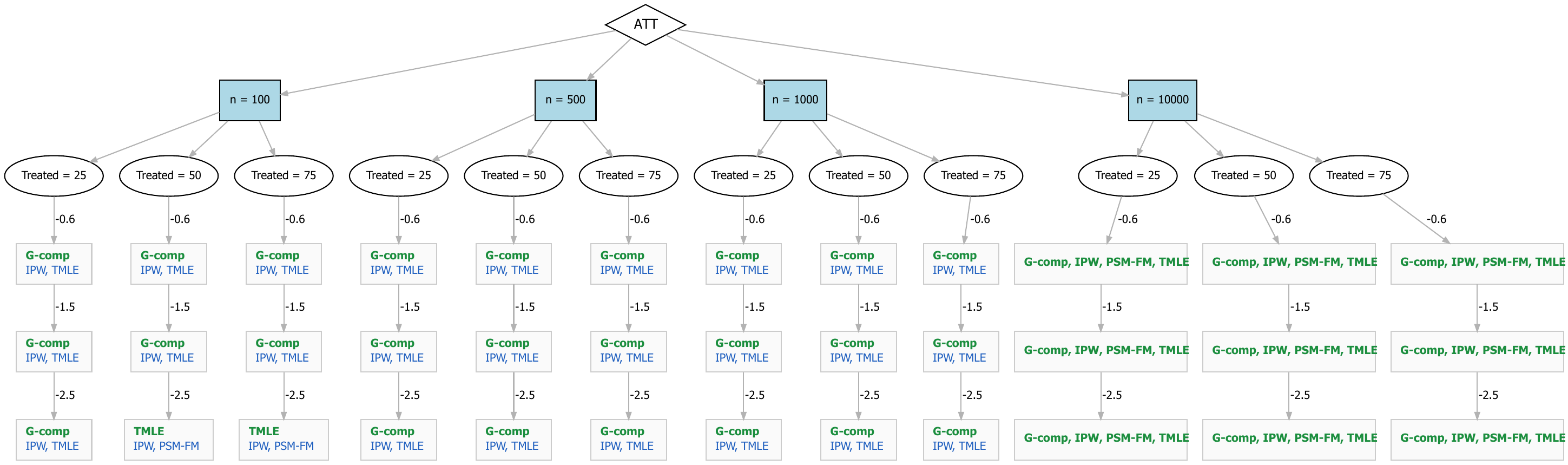}
    \caption{Suggested recommendations from second stage of the simulation. Green indicates the best-performing methods, while blue denotes acceptable but lower-performing alternatives.}  \label{second_recomendation}
  \end{figure}
\end{landscape}
}

\newpage

\section{Applications}\label{sec:4}

We have examined a broad variety of scenarios and have established a comprehensive understanding of how each causal estimation method behaves across this wide range of simulated scenarios. We now turn to the practical question that motivates this work: how should the investigator choose the most appropriate causal methodology when analyzing a real observational dataset?

In this section, we illustrate how the simulation-based guidance developed earlier can be translated into a principled decision framework for empirical applications. Specifically, we take two real clinical datasets and characterize their key design features: sample size, treatment prevalence, outcome prevalence, strength of the treatment effect, and observed PS-OV. We apply the four chosen statistical techniques and give the  marginal OR for the treatment. Once these elements are quantified, we map each dataset to the closest corresponding scenario in our simulation grid. This allows us to compare which statistical technique (IPW, PSM-FM, G-comp, TMLE) is expected to perform best for that particular empirical configuration, based on the previously presented results.

This application shows how simulation-based insights can be used to select the best causal inference tools for real-world clinical research, providing a practical, reproducible approach. Results are presented on the relative scale (marginal OR).

\subsection{COVID-19 study}
First, we analyzed the data published by Rodriguez-Leal et al. \cite{cristobal2025}. Earlier prescriptions of antivirals for mild cases of SARS-CoV-2 infection in vulnerable patients have been shown to lead to better clinical outcomes. However, the specialty of the prescribing physician in the emergency department can affect how long it takes to issue a prescription. Therefore, a total of 534 patients were classified as having received antiviral treatment before or after the third day from symptom onset. The objective of this study is to estimate the causal effect of receiving a prescription from an emergency physician or a specialist physician in another area on receiving treatment after the third day from symptom onset. The confounding variables were immunosuppression, sex at birth, the Charlson index, the type of hospital attending to antiviral prescription frequency, haemodialysis, and microbiological diagnosis by polymerase chain reaction alone. Key characteristics of the sample and results can be found in Table \ref{real_studies}, and Supplementary Section S.5, Figure S.65 (top panel).

As can be seen, for ATE all methods give similar point estimations, but PSM-FM results in the widest CI. The other methods have similar results and performance. For ATT, methods based only in PS (IPW and PSM-FM) and TMLE result in higher point estimations compared to G-comp, and PSM-FM again has the widest CI. Neither of them achieved statistical significance, as all 95\% CIs contained the value of one. The recommendation given in Figure~\ref{second_recomendation} for studies with sample sizes between 500 and 1000 is that PSM-FM should be avoided, which is appropriate for this setting according to the results shown in Table~\ref{real_studies}, both for ATE and ATT.

\subsection{Colorectal surgery study}
Second, to illustrate the selection strategy in another setting, we analyze a multi-center observational dataset including patients who underwent colorectal surgery. The objective is to estimate the causal effect of avoiding a prophylactic drainage (treatment) on the incidence of post-surgical complications (outcome) \cite{blanco2025propensity}, through a database with 3635 patients. This clinical question is particularly relevant given recent conflicting evidence suggesting that drainage, rather than preventing complications, may be associated with risk of postoperative bad outcomes. To rigorously control for potential confounding, the analysis incorporated a comprehensive set of baseline covariates, including age, sex, specific comorbidities (heart disease, diabetes, and hypertension), ASA level, and the type of surgical procedure. Table~\ref{real_studies} shows the distributional properties of the dataset to map it to our simulation scenarios, and the results, also displayed in Supplementary Figure S.65 (bottom panel). The PS-OV between groups is almost 70\%, that is higher than in the previous application. 

For high PS-OV, all methods perform well for ATE and ATT, as shown in Figure~\ref{first_recomendation}. For moderate PS-OV and sample size around 1000, for computing the ATE the recommended techniques are IPW and TMLE, and for the ATT, G-comp, even though TMLE remains a valid alternative, while for a larger sample size, all methodologies perform well. So this setting is allocated in an intermediate place between these sample sizes. Also, there is less heterogeneity in the treatment effect than in the previous application. For ATE, PSM-FM underestimates protective effect and also gives the widest CI, as it is stated in Figure~\ref{second_recomendation}, but for ATT, all methods yielded similar results, as it is expected for a study with a high PS-OV.

\begin{table}[ht]
\centering

\caption{Top rows display the empirical values for the key metrics varied during the simulation study. Middle rows show estimated OR and CI widths for ATE and ATT effects across two real-world studies. Bottom row indicates the recommended technique for each scenario based on the analysis.}
\resizebox{\textwidth}{!}{
\begin{tabular}{lcccccccc}
\toprule
 & \multicolumn{4}{c}{COVID-19 study} & \multicolumn{4}{c}{Colorectal surgery study} \\
\midrule
Sample size & \multicolumn{4}{l}{534} & \multicolumn{4}{l}{3635} \\
Proportion treated & \multicolumn{4}{l}{0.635} & \multicolumn{4}{l}{0.228} \\
Outcome prevalence & \multicolumn{4}{l}{0.403} & \multicolumn{4}{l}{0.258} \\
PS-OV & \multicolumn{4}{l}{0.6385} & \multicolumn{4}{l}{0.696} \\
Conditional log(OR) & \multicolumn{4}{l}{$-$0.746} & \multicolumn{4}{l}{$-$0.673} \\
\midrule
 & \multicolumn{2}{c}{ATE} & \multicolumn{2}{c}{ATT} & \multicolumn{2}{c}{ATE} & \multicolumn{2}{c}{ATT} \\
\cmidrule(lr){2-3} \cmidrule(lr){4-5} \cmidrule(lr){6-7} \cmidrule(lr){8-9}
 & OR & CI width & OR & CI width & OR & CI width & OR & CI width \\
\midrule
IPW & 0.633 & 0.518 & 0.797 & 0.745 & 0.528 & 0.252 & 0.511 & 0.216 \\
PSM-FM & 0.616 & 0.735 & 0.729 & 1.140 & 0.655 & 0.462 & 0.500 & 0.236 \\
G-comp & 0.596 & 0.487 & 0.596 & 0.708 & 0.525 & 0.247 & 0.516 & 0.216 \\
TMLE & 0.647 & 0.551 & 0.816 & 0.738 & 0.525 & 0.252 & 0.511 & 0.265 \\
\midrule
\multicolumn{9}{c}{Recommendation} \\
\midrule
Recommended technique & \multicolumn{2}{c}{IPW and TMLE} & \multicolumn{2}{c}{G-comp} & \multicolumn{2}{c}{IPW and TMLE} & \multicolumn{2}{c}{G-comp} \\
\bottomrule
\end{tabular}%
}
\label{real_studies}
\end{table}

\section{Discussion and future work}\label{sec:5}

\paragraph{Discussion}

The main contribution of this study is a systematic framework for selecting the most appropriate causal inference estimator given the characteristics of the data. Beyond comparing algorithms, we have successfully identified clear performance patterns that correlate directly with key observable data characteristics, specifically, the degree of positivity assumption violation, strength of the effect of the treatment, sample size, proportion of treated patients and prevalence of the outcome. To obtain this we have a designed a wide collection of scenarios. This provides applied investigators with practical selection guidelines: by observing the key characteristics of their real-world observational data, they can now map their scenario to our findings and select the methodology that minimizes bias and improves the estimate for their specific context. Furthermore, the findings from the first and second stages of the simulation complement each other and help to select the most appropriate statistical technique for causal inference in a wide range of scenarios.  

We have also implemented a data-generating mechanism that enables ATE and ATT to differ, meaning that our recommendations can be applied not only to ATE, but also to ATT computations. The need for separate recommendations for the ATE and the ATT is highlighted by the heterogeneity of treatment effects observed in our case studies, more broadly, as is typical in real-world data.

Our results suggest that, while G-comp is generally robust for estimating the ATT, it performs poorly in scenarios involving small sample sizes and low-prevalence outcomes (mild disease). Crucially, our analysis offers a practical solution to this limitation: in these specific rare outcome/small sample configurations, TMLE appear to be a more adequate and stable alternative. For ATE, it is important to highlight that G-comp is biased and overly precise in scenarios with low PS-OV, as this can create a false sense of confidence in inexperienced investigators.

Prior comparative studies have examined this broader array of estimators, yet with different objectives or analytical approaches. For instance, Austin (2007) and Yan Li et al. (2021) evaluated various PS methods focusing on ORs and balancing constraints, respectively; while Cenzer et al. (2020) assessed matching performance specifically in rare disease settings \cite{austin2007, li2021, cenzer2020}. Our research particularly parallels the work by Chatton et al. (2020) who compared G-comp, IPW, and TMLE, but with a primary focus on the impact of different covariate sets \cite{chatton2020}. However, it does not cover the wide and comprehensive range of epidemiological scenarios simulated in our work. Despite this difference in scope, our findings strongly align with their core conclusions: consistent with Chatton et al., we observed that G-comp emerges as the most robust and best-performing methodology across a large portion of the simulated scenarios, especially for ATT, although it also performs well for ATE, particularly in situations where the sample size is not very large and outperforming the IPW and the TMLE in some scenarios with very small sample sizes. Furthermore, our simulation results reveal, in line with Chatton's et al. findings, that PSM-FM consistently yields the poorest performance among the evaluated techniques in the vast majority of cases. In addition, our results demonstrate that IPW and TMLE achieve the best overall performance for ATE across the majority of our simulated scenarios, where the model is consistently well-specified, and G-comp performs well, although not as well as the previous two. If the model is not well-specified, Chatton et al. concluded that G-comp was the most robust methodology against model misspecification, even surpassing TMLE. Our work improves upon these foundational studies by isolating positivity violation, treatment and outcome prevalence and sample size as the critical determinant of performance. Rather than focusing on covariate selection or specific balancing constraints, we systematically mapped how these methods degrade under realistic conditions. 

Our findings regarding estimator performance in restrictive settings respect to sample size are consistent with recent simulation based work and further extend it. Considerable attention has historically been paid to the performance of PSM in small sample contexts. Extensive works by Andrillon et al. (2020), Wan et al. (2019), Bottigliengo et al. (2021), and Pirracchio et al. (2012) have rigorously evaluated specific matching strategies, ranging from replacement techniques to matched versus unmatched analyses and marginal OR estimation\cite{andrillon2020, wan2019, bottigliengo2021, pirracchio2012}. However, our study significantly expands this methodological scope beyond the confines of matching algorithms, incorporating a broader spectrum of modern G-methods (including IPW, G-comp, and TMLE) to provide a more comprehensive overview of the causal inference tools available to epidemiologists.

Although Léger et al. (2022) previously highlighted the critical nature of estimator selection under such positivity constraints\cite{leger2022}, our analysis moves beyond this general warning to quantify performance and identify distinct patterns of choice. By identifying these distinct performance patterns, our study directly addresses the current lack of consensus and the absence of a standardized selection framework noted in very recent literature\cite{burgos2025}, providing the ``diagnostic road-map" necessary to resolve the selection dilemma facing applied researchers. Our recommendations can help researchers to avoid statistical techniques that can give biased or inefficient estimates based on the key characteristics of each database.

In scenarios where positivity violations are extreme (low PS-OV), researchers often consider alternative weighting strategies to stabilize estimation, such as trimmed weights, truncated weights, overlap weights (that compute the average treatment for the overlap population, ATO) and entropy balancing. While these methods successfully reduce variance of the estimator, they come with a significant trade-off: they fundamentally alter the target estimand. By down weighting or discarding units in the tails of the propensity distribution, they no longer estimate the ATE or ATT for the original population, but rather a local effect for a subpopulation with sufficient PS-OV for those patients for whom there is greatest equipoise about treatment selection, like the ATO\cite{Austin2023,Zhou2020}. Therefore, for researchers committed to preserving the original causal estimand (ATE/ATT), our framework prioritizing robust implementations of G-methods remains the recommended approach.

If the investigators are primarily interested in marginal OR, they should be aware of the lack of convergence in some extreme scenarios for IPW and TMLE, mainly when the treatment effect is strong and the sample size is small with low PS-OV. The percentage of no convergence in those settings is higher for IPW and can reach more than 7.0\%. With a higher PS-OV, all methods may also fail to produce results when the outcome frequency and sample size are is low. The percentage is the same for all tested statistical techniques and is around 8.4\% to 15.7\%. Providing information on the failure rate of statistical techniques is also a recommended good practice\cite{morris2019}.

In the COVID-19 dataset, we observed significant heterogeneity (ATE $\neq$ ATT), validating the need for specific estimators for each population. Conversely, in the Surgery dataset, the effect was homogeneous (ATE $\approx$ ATT); in such cases, recommendations derived for the ATE can be reasonably extended to the ATT\cite{Wang2017}.

In conclusion, we have developed a useful handbook through simulation to guide applied researchers in choosing the most appropriate statistical technique when designing observational studies or pragmatic clinical trials that aim to address the causal effect of a binary treatment.

\paragraph{Limitations and future work}

Subsequent studies should replicate this framework under conditions of model misspecification. In the current analysis, we assumed correct model specification to isolate the impact of positivity violations caused by a low overlap of baseline confounders. However, real-world medical data rarely adhere to known functional forms. We hypothesize that under such conditions, the performance hierarchy could shift. TMLE is expected to gain substantial prominence over parametric methods (IPW, G-comp) due to its double-robust property and its ability to integrate machine learning algorithms (via super learner). While parametric TMLE performs well, its resilience is maximized when coupled with data-adaptive learning to mitigate misspecification\cite{TMLE,TMLE_foundation,Smith2022}. Additionally, future work should explore the implementation of bootstrap CIs, which may offer improved precision and coverage stability compared to standard asymptotic approximations in finite samples with positivity issues\cite{Greifer2025}. Other potential areas of research include treatments involving three or more categories, survival models, and studies addressing time-dependent confounding factors.

Due to the factorial design of our study, some potential combinations of key aspects were not explored. However, the first stage of the simulation produced an expected general pattern\cite{Whatif}, prompting us to delve deeper into a common setting involving moderate PS-OV and a moderate-strength treatment, where we have concluded that the proportion of treated patients and  the prevalence of the outcome are only important with small samples ($n=100$). By prioritizing these practically relevant regions of the distribution rather than covering every theoretical permutation, we broadened the scope of our research to provide practical guidance for the specific scenarios that applied investigators are most likely to encounter in real-world data.

Similarly, given the contrast observed in our real-world case studies, where the COVID-19 data exhibited significant treatment effect heterogeneity (ATE $\neq$ ATT) while the Surgery data did not, future simulation designs should systematically model varying degrees of effect heterogeneity. Testing how this heterogeneity interacts with positivity violations will be crucial for refining the specific recommendations for ATE versus ATT estimators under high-variance conditions.

Finally, while our analysis focused on binary outcomes (a prevalent scenario in medical research, e.g., mortality, remission) the methodological architecture developed in this document (including the experimental design, scenario definition, superpopulation generation, and performance metrics) remains valid and easily adaptable for studies with continuous outcomes. Clinical endpoints such as blood glucose levels, oxygen saturation, or estrogen concentrations can be analyzed using this same simulation framework. The primary adjustment required would be the adaptation of the estimands to the distribution of the response variable (e.g., shifting from risk difference to mean difference), but the comparative logic regarding how estimators handle positivity violations would likely hold parallel implications.

\section*{Acknowledgements}\label{sec:6}

AA, JCM, and JA were partially supported by MICIU/AEI/10.13039/501100011033/FEDER,UE [Grants PID2023-150234NB-I00 and PID2024-155426OB-I00]; the Government of Aragon [Research Group E46\_23R: Modelos Estocásticos]; and INFORMA-SL [Grant UZ-2026/0114].

CMRL and RS were partially supported by MICIU/AEI/10.13039/501100011033/FEDER,UE [Grant PID2024-155289NB-I00]; the Community of Madrid, 2025 Expenditure Budget [Fundaciones de Investigación Biomédica]; and Fundación Española de Medicina de Urgencias y Emergencias (FEMUE) [FEMUE doctoral thesis support].

TP was partially supported by MICIU/AEI/10.13039/501100011033/FEDER,UE [Grant PID2022-137050NB-I00].

All authors are members of the BIOSTATNET network [RED2024-153680-T], funded by MICIU/AEI/10.13039/501100011033.

\bibliographystyle{wileyNJD-AMA}
\bibliography{referencias}

\end{document}